\documentclass[prd,preprint,a4paper,superscriptaddress,nofootinbib,11pt]{revtex4}
\usepackage{amsmath,amsfonts,amssymb}
\usepackage{txfonts}
\usepackage[T1]{fontenc}
\usepackage{graphicx}
\graphicspath{ {images/} }

\begin{document}


\begin{center}
{\bf  \Large Could LIV/noncommutativity alleviate  the cosmological constant problem  }

\bigskip
\bigskip

   Andjelo Samsarov {\footnote{e-mail: asamsarov@irb.hr }}\\

 {\it Rudjer Bo\v skovi\' c Institute, Theoretical Physics Division\\
Bijeni\v cka 54, 10002 Zagreb, Croatia}\\


\end{center}
\setcounter{page}{1}
\bigskip


\begin{center}
{\bf   Abstract}
\end{center}

    We test the hypothesis that in models with an  explicit breaking of Lorentz invariance (LIV models) the problem with the cosmological constant
 may, if not already be solved, then at least may  partially be  mitigated by  improving its UV behaviour. Certain class of noncommutative models that have  a well defined dual momentum space geometry  and in addition set up a suitable frame for developing the appropriate  DSR  models
are also subject to the same test.

\newpage
\section{Introduction}

A major outstanding problem in theoretical physics is that most quantum field theories predict a huge value for the energy of the quantum vacuum.
 On the other hand a common assumption is that the  cosmological constant is equivalent to the energy of the quantum vacuum (ground state enery of the quantum field).
  Although no theory exists that supports this assumption, arguments can be made in its favor \cite{zeldovich}, \cite{zeldovich1}.
Such arguments are usually based on dimensional analysis and effective field theory.
 If the universe is described by an effective local quantum field theory down to the Planck scale,
  then we would expect a cosmological constant of the order of $M_{ Pl}^4$.
  However, the problem is that the actual measured cosmological constant is smaller than this by a factor of $10^{�−120}$.
  This discrepancy has been called 
  the cosmological constant problem, by all accounts the worst theoretical prediction in the history of physics.
  The additional problem is that so far
   there is no known natural way to derive the tiny cosmological constant used in cosmology from particle physics.

On the other hand  different approaches to quantum gravity suggest that Lorentz symmetry might be violated at very high energies. They not only propound a possible Lorentz invariance violation (LIV), but also call for the rethinking of  one's notion
of a differential manifold, as used to describe a spacetime. In particular this refers to questioning  the validity of a usual  description of spacetime as a smooth differential manifold.
Moreover, lacking a full theory of quantum gravity, one turns to effective field theories whose main scope is to bring about
a sufficiently convenient account of physical phenomena at accordingly  lower scales, with a  hope   that  these theories may capture
some of the more representative features of the full  theory of quantum gravity.
One feature that might be considered as sufficiently representative of quantum gravity effects
would be an appearance 
of some specific signal of   Lorentz  symmetry breaking. In this sense Lorentz invariance  violation may be regarded as one of the more striking remnants of quantum gravity at an accordingly lower energy scales.



Among  different effective field and   phenomenological  approaches to quantum gravity, we focus
on two among them which in the present context may be considered as the most interesting ones: the Lorentz invariance violation (LIV) models \cite{Jacobson:2005bg},\cite{Colladay:1998fq},\cite{Myers:2003fd}
and the  doubly (or deformed) special relativity models  (DSR)
\cite{AmelinoCamelia:2000mn,AmelinoCamelia:2000ge,Magueijo:2001cr,Magueijo:2002am,KowalskiGlikman:2004qa,Borowiec:2010yw}.
Unlike DSR models, in LIV models the observer independence is explicitly violated and a preffered frame
is singled out. Preffered frame in this case is typically thought to be the rest frame
of the cosmic microwave background radiation. This means giving up of the relativity principle and the equivalence of all inertial observers.

On the other side, in DSR models the
postulates of relativity may be reformulated in such a way as to evade a necessity for singling out
one preffered frame. This requires an introduction of one additional invariant scale (the mass scale $\kappa$), besides that of the speed of light $c,$ and modifying the 	Lorentz transformations accordingly. 
 DSR models are quite often realised within a framework of noncommutative (NC) spaces
and they provide a description of the kinematical aspects of the effective, quantum gravity derived theories   at accordingly lower scales. Although the invariant  energy\footnote{Throughout the paper the analysis is carried in units $\hbar = c =1.$}  
scale $\kappa$
 is often taken to be the Planck mass $M_{Pl},$ this doesn't necessarily need to be so. Indeed, in some cases it is natural to have $\kappa$ getting closer to the values as low as the typical energies of the ultra high energy cosmic rays (UHECR).

In this article we examine a possibility that the  Lorentz invariance violation and/or the  modified structure of spacetime (both emerging at some scale $\kappa$) might, if not solve, then at least mitigate the problem with the cosmological constant by reducing to a certain extent a huge discrepancy  between its measured and the theoretically predicted value. Essentially, the  idea is that  due to the Lorentz invariance violation on the one side or the noncommutative nature of spacetime on the other,  the relevant scalar field operator gets deformed in such a way as to be able to exhibit a somewhat
 improved  behaviour in the UV regime.
We test this hypothesis within the frameworks of two phenomenological  theories mentioned before,  LIV and DSR,  respectively. The cosmological constant within a  noncommutative setting was also studied in 
\cite{garatini},\cite{Garattini:2011kp},\cite{RezaeiAghdam:2004zh},\cite{Calmet:2005mc}.
Likewise, the properties of the vacuum energy   have  been analysed  for a field theory
on  NC spaces with different topologies,
  showing that the vacuum energy may  become finite if 
  some  restrictions
on  the type of NC space  were imposed \cite{Chaichian:2001pw}. 
The finite result for the vacuum energy has also been obtained by considering the  graviton contribution to the induced cosmological constant at one loop \cite{garatini},\cite{Garattini:2011kp}.
 The analysis in this paper is carried in an arbitrary number of dimensions $d$ whenever possible, and each time when  it was reduced  to some particular $d,$ it was explicitly emphasized.


\section{The field operator and the corresponding heat kernel trace}

Suppose that a bosonic quantum  field theory
  in a Euclidean flat spacetime is  described by the field
   operator $F(\Box),$  where $\Box = - \partial_{\mu}
   \partial^{\mu}$ is the usual d'Alembertian operator written in the  Euclidean metric. This means  that the Klein-Gordon equation
   governing the dynamics of the field is
\begin{equation}
  (F(\Box) + m^2 ) \Phi = 0,
\end{equation}
for a field of mass $m$. The corresponding free field propagator
in momentum space is then given by
\begin{equation}
 G(p^2) = \frac{1}{F(p^2) + m^2},
\end{equation}
as a generalization of the standard propagator $1/(p^2 + m^2)$ valid
in the theory described by the $4$-dim Laplace operator $\Box = - \partial_{\mu}
   \partial^{\mu}$.

The generating functional (i.e. the partition function) of the theory is then given by
\begin{equation}
 Z = \int {\mathcal{D}} \Phi ~ e^{-S} = \int {\mathcal{D}} \Phi ~
 e^{\int d^d x \Phi (x)
 (F(\Box) + m^2 ) \Phi (x)} = \frac{1}{\sqrt{\det (F(\Box) + m^2 )}}.
\end{equation}
The effective action then reads
\begin{equation} \label{effectiveaction}
  W_{eff} = \ln Z = \ln {(\det (F(\Box) + m^2 ))}^{-\frac{1}{2}} =
  -\frac{1}{2}
 \int_{\epsilon^2}^{\infty} \frac{ds}{s} {\mbox{Tr}} ~ e^{-s(F(\Box) + m^2 ) },
\end{equation}
where $\epsilon$  is a UV cutoff,  $~ \epsilon << 1, ~$ introduced at the lower limit in the above integral since in the standard theory it usually diverges. It is worthy to mention that the heat kernel trace \cite{vassilevich} that enters the equation (\ref{effectiveaction})  is also of essential importance for studying the spectral dimension \cite{Ambjorn:2005db}.

In the next section we present some of the most well-known models with deformed dispersion
 relations in Minkowski space with signature $(-,+,+,+)$. They all fit within a DSR framework and may be viewed as being derived from a set 
of noncommutative  spaces, modelling a noncommutative nature of spacetime.   The deformation parameter $\kappa$ that appears in these models and
that sets a noncommutative scale has the dimension of mass and is often identified
with the Planck mass $\kappa \sim M_{Pl}$, although it is not mandatory. The models can be characterized from the
Casimir operator $F(p)$ which gives rise to the dispersion relation $F(p)=-m^2$.

\section{ Vacuum density in spaces modelling noncommutative nature of spacetime}

  The first model that we analyse  is the noncommutative model related to Snyder space \cite{Snyder:1946qz}.
  The symmetry of this space is characterized by the Casimir operator which may be an arbitrary function of $p^2$ since its form is not unique. However, one usually takes the following   form  
\begin{equation} \label{Snyderbasis}
 F(p) = \frac{p^2}{1 - p^2/\kappa^2},
\end{equation}
 where $p^2=-p_0^2+p_i^2$.

    Another interesing candidate for describing the noncommutative features of spacetime is
  the Majid-Ruegg model \cite{Majid:1994cy,AmelinoCamelia:2000mn,AmelinoCamelia:2000ge}, which
is  described by the Casimir
\begin{equation} \label{bicrossproductbasis}
 F(p) = - {\bigg( 2\kappa \sinh\left(\frac{p_0}{2\kappa}\right) \bigg)}^2 + {\bf{p}}^2 e^{\frac{p_0}{\kappa}}.
\end{equation}
The third one that is considered here is the  Magueijo-Smolin model \cite{Magueijo:2001cr,Magueijo:2002am}, with the dispersion
\begin{equation} \label{maguejosmolinbasis}
  F(p) = \frac{-p_0^2 + {\bf{p}}^2 }{ {\bigg(1 - \frac{p_0}{\kappa} \bigg)}^2 }.
\end{equation}
 As well as the Majid-Ruegg model, it is often connected to  $\kappa $-Minkowski noncommutative spacetime. As a matter of fact, both these models can be seen as  different 
bases \cite{KowalskiGlikman:2002we,KowalskiGlikman:2002jr}
 of the $\kappa-$Poincar\'{e} algebra \cite{Lukierski:1991pn}, which is the symmetry algebra of the  $\kappa $-Minkowski spacetime.
Though these two models give representations of the same NC space,
   one pertaining to the  bicrossproduct or Majid-Ruegg basis
   \cite{Majid:1994cy,AmelinoCamelia:2000mn,AmelinoCamelia:2000ge}, 
 and the other referring to Magueijo-Smolin basis, they generally lead to different physics. 

As suggested from the beginning,
the formalism with the heat kernel is adapted to the Euclidean setting.
Therefore,
in order to proceed further, we switch to the Euclidean notation, which
corresponds to adopting the new variables
\begin{equation} \label{wickrotation}
  p_d = i p_0,
\end{equation}
where $p^{Lorentz} = (p_0, p_1,...,p_{d-1})$  and  $p^{Euclid} =
(p_1,...,p_{d-1}, p_d)$ are the momentum $d$-vectors written in
Lorentzian and Euclidean notations, respectively.  Also, an important point here   is to  establish  a behaviour of the constant $\kappa$ under Euclideanization. As far as this matter goes, the constant $\kappa$ is kept intact throughout the whole analysis carried here. However, for other approaches to the Euclideanization of NC spaces see \cite{euclideanization,benedetti,Arzano:2014jfa}.  
 It is important to  note here that the heat kernel method requires the use of
the Euclidean version of the theory since otherwise the path integral  (3) could  not
be properly defined. Besides that, the evaluation of the heat kernel trace requires
integration over the variables that  have been analytically continued previously.
 Consequently, all of them will have been integrated out
in the process of taking the trace. Finally, the resulting trace will depend only
on the small adiabatic parameter $s$ and the mass scale $\kappa$ describing the deformation.
 Upon  changing
the variables and expanding up to the first order in $\frac{1}{\kappa},$
one arrives at the different linearized forms for the Casimirs, which we
analyse in a consecutive order.

\subsection{Vacuum density for the Snyder model}

In the case of Euclidean Snyder space the Casimir, after expanded, takes the form
\begin{equation} \label{Snyderbasis-expand}
 F(p) = p^2 (1 +  \frac{1}{\kappa^2} p^2 ),
\end{equation}
that is
\begin{equation} \label{bicrossoperator}
  \hat{O} \equiv F( \Box) = p^2  +  \frac{1}{\kappa^2}  p^4   =  \Box \bigg( 1  +   \frac{1}{\kappa^2}   \Box  \bigg),
\end{equation}
where $ \Box = -\partial_{\mu } \partial_{\mu }. $
This is the Lorentz invariant theory. The heat kernel in this case is 
known \cite{Nesterov:2010yi} and reads
\begin{equation} \label{traceheatkernelsnyder}
   \mbox{Tr} \; e^{-s F( \Box ) } \; = \; \frac{1}{{(4\pi)}^{d/2}} V
   P_d (s),  \quad  P_d (s) =  \frac{2}{\Gamma (\frac{d}{2})}
   \int_0^{\infty} dp \; p^{d-1} e^{-s F(p^2)}.
\end{equation}
Using the result (\ref{effectiveaction}) for the effective action, $W_{eff} =
-\Lambda (\epsilon) V,$ gives  rise to  the vacuum energy density
\begin{equation} \label{vac_snyder_1}
   \Lambda (\epsilon ) \; = \; \frac{1}{2} \frac{1}{{(4\pi)}^{d/2}}
  \int_{\epsilon^2 }^{\infty} \frac{ds}{s} P_d (s).
\end{equation}

In $d=4$ dimensions, a straightforward calculation leads to
\begin{equation} \label{vac_snyder_2}
   \Lambda (\epsilon ) \; = \; \frac{1}{4} \frac{ \kappa^2}{{(4\pi)}^{d/2}}
  \int_{\epsilon^2 }^{\infty} \frac{ds}{s^2} \bigg( 1+ \frac{
  \kappa \sqrt{\pi} }{2  } \sqrt{s} \; e^{\frac{s  \kappa^2}{4 }}
 \bigg( \mbox{erf}  (\frac{\kappa \sqrt{s}}{2  }) -1 \bigg)  \bigg),
\end{equation}
where $\mbox{erf} (x) = \frac{2}{\sqrt{\pi}} \int_0^x dz \; e^{-z^2}$  is the
Gauss error function. Focusing on the  regime  $s  \kappa^2   >> 1,$  which is the only one that is compatible with a low energy limit in which the inverse propagator (\ref{Snyderbasis-expand}) has been derived, one finds 
 the behaviour of the function $P_d(s) \sim s^{-2}.$ Consequently, this leads to the vacuum energy density
of the form $\sim \frac{1}{\epsilon^4},$ which therefore
 exhibits no UV improvement when compared to the vacuum energy resulting from the standard field operator.

\subsection{Vacuum density for the Majid-Ruegg model}

In order to analyse a behaviour of the cosmological constant in the
noncommutative space with $\kappa$-type of deformation, which entails a certain DSR framework, we continue to
work  with a Euclidean notation and thus we again switch to the variables
\begin{equation} \label{wickrotation}
  p_d = i p_0, \quad  M=  -im,
\end{equation}
where $p^{Lorentz} = (p_0, p_1,...,p_{d-1})$  and  $p^{Euclid} =
(p_1,...,p_{d-1}, p_d)$ are the momentum $d$-vectors written in
a Lorentz and Euclidean notation, respectively. Upon carrying out  the transformation (\ref{wickrotation})
 and expanding  (\ref{bicrossproductbasis})  up to a first order in $\frac{1}{\kappa},$
one arrives at the following form for the Casimir,
\begin{equation} \label{bicrossproductbasis-expand}
  M^2 = {p_d}^2 + {\bf{p}}^2 - \frac{i}{\kappa} {\bf{p}}^2  p_d.
\end{equation}
Similarly, the coordinates undergo the same transformation, $ x_d = i
x_0,$ (meaning $\tau = -it,$) where, as in the case with the momenta, $x^{Lorentz} = (x_0 = -t, x_1,...,x_{d-1})$  and  $x^{Euclid} =
(x_1,...,x_{d-1}, x_d \equiv \tau)$ are the coordinate $d$-vectors written in
a Lorentz and Euclidean notation, respectively.

By the standard prescription, $p_i = -i \frac{\partial}{\partial x_i}$
and $p_d = -i \frac{\partial}{\partial \tau}, $ it is straightforward
to identify the field operator $\hat{O},$ for the theory described by
the Casimir (\ref{bicrossproductbasis-expand})
\begin{equation} \label{bicrossoperator}
  \hat{O} \equiv \Box_{\kappa} = -{\partial_{\tau}}^2 - \partial_i
  \partial_i + \frac{1}{\kappa} \partial_i \partial_i \partial_{\tau}.
\end{equation}
Namely, the latter  follows from the comparison with the Klein-Gordon equation $\Box_{\kappa} - M^2 =0$.
Having that, it is of particular interest  to find the heat kernel for this operator. With this
purpose, one needs to solve the heat diffusion equation
\begin{equation} \label{bicrossoperator-heatkernel}
  \bigg( \frac{\partial}{\partial s} + \hat{O} \bigg) \; K(s, {\bf{x}},
  {\bf{x'}}; \tau, \tau' ) \; = \; 0,
\end{equation}
with the constraint
\begin{equation} \label{constraint}
  K(s=0, {\bf{x}}, {\bf{x'}}; \tau, \tau' ) \; = \; \delta^{(d-1)}
  ({\bf{x}}- {\bf{x,}}) \; \delta (\tau - \tau').
\end{equation}
 Heat diffusion equation on $\kappa-$type spaces has been considered in different contexts and approaches
     in \cite{benedetti},\cite{Arzano:2014jfa},\cite{Anjana:2015ios}.

A solution to the equation  (\ref{bicrossoperator-heatkernel}) can be searched in a form of the Fourier transform
\begin{equation} \label{Fourier}
  K(s, {\bf{x}}, {\bf{x'}}; \tau, \tau' ) \; = \;
  \int \frac{d \omega}{2\pi} \int \frac{d^{d-1}p}{{(2\pi)}^{d-1}}
  \; e^{i\omega (\tau - \tau')}  e^{i{\bf{p}} \cdot  ({\bf{x}} -
  {\bf{x'}})} U({\bf{p}}, \omega, s).
\end{equation}
By plugging this into equation (\ref{bicrossoperator-heatkernel}) with the operator $\hat{O}$ given by
(\ref{bicrossoperator}), one gets $ \; U({\bf{p}}, \omega, s) = e^{s(- \omega^2 -{\bf{p}}^2 + i \frac{\omega}{\kappa} {\bf{p}}^2)}, \;$ leading to
\begin{equation} \label{bicrossoperator-heatkernel1}
  K(s, {\bf{x}}, {\bf{x'}}; \tau, \tau' ) \; = \; \int \frac{d \omega}{2\pi} \int \frac{d^{d-1}p}{{(2\pi)}^{d-1}}
    \; e^{i\omega (\tau - \tau')}  e^{i{\bf{p}} \cdot  ({\bf{x}} -
    {\bf{x'}})} \;  e^{-s( \omega^2 +{\bf{p}}^2 - i \frac{\omega}{\kappa} {\bf{p}}^2)} \; \equiv \; \langle {\bf{x}}, \tau | e^{-s \Box_{\kappa}} | {\bf{x'}}, \tau' \rangle.
\end{equation}
Having this, it is straightforward to calculate the trace of the heat kernel as
\begin{eqnarray} \label{bicrossoperator-traceheatkernel}
 \mbox{Tr} \; K(s) \; & = &\; \mbox{Tr} \; e^{-s \Box_{\kappa}} \; = \;
    \int d \tau \int d^{d-1}x \; \langle {\bf{x}}, \tau | e^{-s \Box_{\kappa}} | {\bf{x}}, \tau \rangle   \nonumber \\
    & = & \; \frac{\Delta \tau V}{{(2\pi)}^{d}} \;
       \int  d^{d-1}p \int_{-\infty}^{\infty} d\omega
     \;  e^{-s( \omega^2 +{\bf{p}}^2 - i \frac{\omega}{\kappa} {\bf{p}}^2)}   \\
     & = & \frac{\Delta \tau V}{{(2\pi)}^{d}}  \frac{\sqrt{\pi}}{\sqrt{s}} \int_{0}^{2\pi} d \phi_{d-2} \int_{0}^{\pi} d \phi_{d-3} \cdot \cdot \cdot \int_{0}^{\pi} d \phi_1 \sin^{d-3} \phi_1
       \sin^{d-4} \phi_2  \cdot \cdot \cdot \sin \phi_{d-3} \int_{0}^{\infty} {|{\bf{p}}|}^{d-2} \;
       e^{-s {\bf{p}}^2 - \frac{s}{4 \kappa^{2}} {\bf{p}}^4 } d|\bf{p}|,   \nonumber
\end{eqnarray}
which  gives the following result
\begin{equation} \label{bicrossoperator-traceheatkernel1}
  \mbox{Tr} \; e^{-s \Box_{\kappa}} \; = \; \frac{\Delta \tau V}{{(4\pi)}^{d/2}} \frac{1}{\Gamma(\frac{d-1}{2})} \frac{2}{\sqrt{s}} \int_{0}^{\infty} {|{\bf{p}}|}^{d-2} \;
         e^{-s {\bf{p}}^2 - \frac{s}{4 \kappa^{2}} {\bf{p}}^4 } d|\bf{p}|.
\end{equation}
After performing the integration over the absolute value of the three-momentum, the result is finally
\begin{equation} \label{bicrossoperator-traceheatkernel2}
  \mbox{Tr} \; e^{-s \Box_{\kappa}} \; = \; \frac{\Delta \tau V}{{(4\pi)}^{d/2}} \frac{1}{\Gamma(\frac{d-1}{2})} \frac{2}{\sqrt{s}}
   \bigg[ {(\sqrt{2})}^{d-5} {\bigg(\frac{s}{\kappa^2} \bigg)}^{-\frac{d}{4} - \frac{1}{4}}
  \bigg( \sqrt{\frac{s}{\kappa^2}} \Gamma(\frac{d-1}{4}) M(\frac{d-1}{4}, \frac{1}{2}, s \kappa^{2})
     -2s \Gamma(\frac{d+1}{4})  M(\frac{d+1}{4}, \frac{3}{2}, s \kappa^{2})\bigg) \bigg],
\end{equation}
where $M(a,b,z)$  is the Kummer confluent hypergeometric function defined by the series
\begin{equation} \label{kummer}
  M(a,b,z) = \sum_{n=0}^{\infty} \dfrac{a(a+1) \cdots (a+n-1)}{b(b+1) \cdots (a+n-1)} \dfrac{z^n}{n!}.
\end{equation}

We now take the asymptotic limit $s \kappa^2 >> 1$ which is compatible
 with the limiting case in which the Laplacian operator
 (\ref{bicrossoperator}) is derived. In this limit the heat kernel trace looks as
\begin{eqnarray} \label{bicrossoperator-traceheatkernel-umetak}
  \mbox{Tr} \; e^{-s \Box_{\kappa}} \;  &=& \;  {( s \kappa^{2})}^{-\frac{d}{4}} {\big( \frac{s}{\kappa^2} \big)}^{-\frac{d+1}{4}}
   \bigg[  \frac{2^{-\frac{d+3}{2}}  \pi^{-\frac{d}{2}} \Delta \tau V   }{\Gamma(\frac{d-1}{2})}
    \bigg(  \frac{{(-1)}^{\frac{1-d}{4}}   \frac{\sqrt{\pi}}{\kappa} \Gamma(\frac{d-1}{4}) }{\Gamma(\frac{1}{2}+\frac{1-d}{4})}   
   -   \frac{{(-1)}^{-\frac{1+d}{4}}   \frac{\sqrt{\pi}}{\kappa} \Gamma(\frac{d+1}{4}) }{\Gamma(\frac{3}{2}-\frac{d+1}{4})} \bigg)  
    {(s\kappa^2)}^{\frac{1}{4}}   \nonumber  \\ 
   & +  &  \frac{ {(-1)}^{\frac{3-d}{4}}  \pi^{\frac{1-d}{2}} \Delta \tau V   }{\kappa \Gamma(\frac{3}{2}-\frac{d+1}{4})         
                 \Gamma(\frac{1}{2}+\frac{1-d}{4}) \Gamma(\frac{d-1}{2})   }
               \bigg( i  \Gamma(\frac{3}{2}-\frac{d+1}{4}) \Gamma(\frac{d-1}{4})    \\
& -&   \Gamma(\frac{1}{2}+\frac{1-d}{4}) \Gamma(\frac{1+d}{4}) \bigg) 
\bigg(\frac{2^{-\frac{d+11}{2}} (d^2 - 1)}{ {(s\kappa^2)}^{\frac{3}{4}}} - \frac{2^{-\frac{d+21}{2}} (d^4 + 8d^3 + 14d^2 -  8d  - 15)}{{(s\kappa^2)}^{\frac{7}{4}}} \bigg)  + O \bigg({(\frac{1}{s\kappa^2})}^{\frac{11}{4}} \bigg) \bigg].   \nonumber
\end{eqnarray}
In the above expression we have used the asymptotic expansion for the Kummer's confluent hypergeometric function \cite{abramowitz}.
Since in the limit   $s \kappa^2 >> 1$   all terms are neglected except the first one, it follows that $~ \mbox{Tr} \; e^{-s \Box_{\kappa}}   \sim   s^{-d/2},~$ which is the standard (commutative) behaviour. Correspondingly, $~\Lambda(\epsilon) \sim 1/\epsilon^4, ~$ meaning that there is no improvement in the UV behaviour of the cosmological constant.

\subsection{Vacuum density for the Magueijo-Smolin model}

  In order  to see what happens in the Magueijo-Smolin model,
we analyse a low-energy limit of the Casimir (\ref{maguejosmolinbasis}), which upon Euclideanization takes  the form
\begin{equation} \label{mscasimir}
  -m^2 = M^2 = {p_d}^2 + {\bf{p}}^2 - \frac{2i}{\kappa}
   {\bf{p}}^2 p_d  - \frac{2i}{\kappa}  {p_d}^3 + {\mathcal{O}} ({(1/\kappa)}^2).
\end{equation}
This leads to an appropriate scalar field operator, i.e. the  Laplacian
\begin{equation} \label{lowenergymaguejosmolin}
  \hat{O} \equiv \Box_{\kappa}= -{ \partial_\tau }^2  - \partial_i \partial_i + \frac{2}{\kappa} \partial_i \partial_i  \partial_\tau + \frac{2}{\kappa} {\partial_\tau}^3.
\end{equation}
  Since we want  to find a heat kernel for the operator (\ref{lowenergymaguejosmolin}), we again turn to the equation (\ref{bicrossoperator-heatkernel}), which this time has to be solved
for the field operator (\ref{lowenergymaguejosmolin}). Assuming the solution in a form of the Euclidean Fourier transform (\ref{Fourier}), one gets in this particular case
$$
  U({\bf{p}}, \omega, s) = e^{-s \bigg(\omega^2 + {\bf{p}}^2
 - \frac{2i}{\kappa}
   \omega {\bf{p}}^2  - \frac{2i}{\kappa}  {\omega}^3 \bigg) }
$$
and consequently for the trace of the heat kernel
\begin{eqnarray}
 \mbox{Tr} \; K(s) \; & = &\; \mbox{Tr} \; e^{-s \Box_{\kappa}} \; = \;
    \int d \tau \int d^{d-1}x \; \langle {\bf{x}}, \tau | e^{-s \Box_{\kappa}} | {\bf{x}}, \tau \rangle   \nonumber \\
    & = & \; \frac{\Delta \tau V}{{(2\pi)}^{d}} \;
       \int  d^{d-1}p \int_{-\infty}^{\infty} d\omega
     \;  e^{-s( \omega^2 +{\bf{p}}^2)}
    e^{s \bigg( 2 i \frac{\omega}{\kappa} {\bf{p}}^2 + 2 \frac{i}{\kappa} {\omega}^3 \bigg) }.  
 \label{maguejosmolin-traceheatkernel}
\end{eqnarray}

The integral for the heat kernel trace is rather involved, rendering an exact evaluation extensively difficult to perform.
Despite the  perplexities involved, one may however search to
 find a reliable estimation for the above integral which would eventually  retain
all essential features of the heat kernel trace, enabling us to deduce the actual UV structure of the vacuum energy. 
The details of the calculation are presented in the Appendix A.

Based on the result obtained in the Appendix A, it is possible to write  a required heat kernel trace in the form
\begin{eqnarray}  \label{maguejosmolin-traceheatkernel11}
 \mbox{Tr} \; K(s) \; & = &\; 
     \frac{\Delta \tau V}{2^d \pi^{d/2}} \;
     \frac{1}{\sqrt{\pi}}  \frac{1}{s^{\frac{d-1}{2}}} \int_{-\kappa}^{\kappa} 
    \; \frac{e^{-s \omega^2 } e^{i \frac{2s}{\kappa} {\omega}^3 }}{{\bigg( 1 - \frac{2i}{\kappa} \omega     \bigg)}^{\frac{d-1}{2}}}  d\omega   \nonumber \\
  & = & \; \frac{\Delta \tau V}{2^d \pi^{d/2}} \;
     \frac{1}{\sqrt{\pi}}  \frac{1}{s^{\frac{d-1}{2}}} \Bigg(  - \oint_{\Gamma}
    \; \frac{e^{-s \omega^2 } e^{i \frac{2s}{\kappa} {\omega}^3 }}{{\bigg( 1 - \frac{2i}{\kappa} \omega      \bigg)}^{\frac{d-1}{2}}}  d\omega   + \int_{\gamma (R)}
    \; \frac{e^{-s \omega^2 } e^{i \frac{2s}{\kappa} {\omega}^3 }}{{\bigg( 1 - \frac{2i}{\kappa} \omega     \bigg)}^{\frac{d-1}{2}}}  d\omega  \Bigg),
\end{eqnarray}
which subsequently gives rise to the effective action
\begin{eqnarray}  \label{maguejosmolin-traceheatkernel12}
    W_{eff} [1/ \kappa] \; & = & \;  -\frac{1}{2} \int_{\epsilon^2}^{\infty} \frac{ds}{s} ~\mbox{Tr} \; K(s)  \nonumber \\
   & = & \; -\frac{1}{2} \int_{\epsilon^2}^{\infty} \frac{ds}{s} ~ \frac{\Delta \tau V}{2^d \pi^{d/2}} \;
     \frac{1}{\sqrt{\pi}}  \frac{1}{s^{\frac{d-1}{2}}} \Bigg(  -  2\pi i ~ {\mbox{Res}}_{\omega = -i \frac{\kappa}{2}} f(\omega)  \nonumber \\  
  &+&  \frac{e^{s\kappa^2}}{9^{\frac{d-1}{4}}} \Bigg[  \frac{\pi}{6\kappa s} \bigg(1- e^{-2s \kappa^2}   \bigg)   + \frac{\pi^{3/2}}{3} \frac{e^{2s\kappa^2}}{\sqrt{2s}} {\mbox{erf}} \bigg(\sqrt{2s \kappa^2} \bigg) \Bigg]  \Bigg),
\end{eqnarray}
with $~f(\omega) ~$  being displayed in (\ref{appendix4}).

 On the basis of the result just obtained it is straightforward to deduce on the UV behaviour of the vacuum energy
in spaces whose dimensionality is an odd number.  As before, it is necessary to consider a limit which is compatible with that
in which the operator (\ref{lowenergymaguejosmolin}) has been derived. With $s \kappa^2 >> 1$ being the right limit, we find the relevant trace of the heat kernel as
\begin{eqnarray} \label{appendix13-dodatak}
 \mbox{Tr} \; K(s) \;  = \;    \frac{\Delta \tau V}{2^d \pi^{d/2}} \;
     \frac{1}{\sqrt{\pi}}  \frac{1}{s^{\frac{d-1}{2}}} \Bigg(  -  2\pi i ~ {\mbox{Res}}_{\omega = -i \frac{\kappa}{2}} f(\omega)  
  +  \frac{\pi}{3^{\frac{d+1}{2}}} \Bigg[  \frac{\sinh s \kappa^2}{s \kappa}
   + \sqrt{\pi} \frac{e^{3s\kappa^2}}{\sqrt{2s}} \Bigg(  1-  \frac{1}{\sqrt{\pi}}   \frac{e^{-2s\kappa^2}}{\sqrt{2s \kappa^2}}      
    + O(\frac{1}{s\kappa^2})     \Bigg)    \Bigg] \Bigg).   \nonumber 
\end{eqnarray}
In the last expression  the asymptotic expansion for the error function was used. Retaining the most dominant terms, $\mbox{Tr} \; K(s)$
 takes the form $~ \mbox{Tr} \; K(s)  \sim   \alpha_d \frac{ e^{3s\kappa^2}}{s^{d/2}} + \frac{P_d (s)}{s^{(d-1)/2}}, ~$
with $P_d (s)$ and $\alpha_d$ being some polynomial in $s$ and a constant, respectively, both depending on $d$. It is clear from here that nothing is going to change  in the UV regime, except only that the standard $1/\epsilon^4$ divergency will  get smeared
a bit  by the factor $e^{3s\kappa^2}.$  In the IR regime instead, the situation gets seriously aggravated with respect to the standard field theory.

  Few comments are in order. To begin with, it may be noted that in the three models analysed so far, each representing a different DSR proposal and each with
an appropriate underlying noncommutative substructure, the UV cutoff $~\epsilon \sim 1/M_{UV}~$ needs to be such that the condition $~M_{UV} << \kappa ~$ holds true. The reason for this is  that we don't consider the exact field operators of the above studied models, but only their approximations to a first order in $1/\kappa$ and on top of that, these approximations are valid only if the latter requirement is fulfilled. This however poses no problems provided the NC scale $\kappa$ has been fixed at the Planck scale, namely,  $~\kappa \sim M_{Pl} ~$.  Actually, in  view of some recently propelled proposals, such as the weak gravity conjecture \cite{arkanihamed,Huang1,Huang2,Huang3}, the UV cutoff of the theory may be suppressed well below the Planck scale, making  
the above analysis fit into that scheme.

   Another point to note
 is that all three DSR models studied here  rely on certain symmetries,  be they  deformed or not, which underlie the   noncommutative  spacetime settings used to formulate these models.
By way of, they all exhibit certain symmetry features that are revealed  especially  through a geometric
character of their 
corresponding dual momentum spaces. For instance, the $\kappa-$Minkowski space, which may be viewed as a suitable arena for devising DSR models such as the Majid-Ruegg model or Magueijo-Smolin model,  is known to have a dual 
curved momentum space, which encompasses  one   half of  de Sitter space \cite{KowalskiGlikman:2002ft},\cite{KowalskiGlikman:2004tz}. Incidently, the latter 
 emerges as a group manifold of the Lie group 
 AN(d-1) (for the theory in $d$ dimensions), obtained by exponentiating the $\kappa-$Minkowski algebra and by identifying the group parameters with the local momentum coordinates.

   The existence of this additional geometric and symmetry infrastructure may likely be utilised  for creating a  framework 
   that could be more suitable for carrying the  analysis of the type studied here and could  eventually  make for the calculations become more accurate. 
In particular, the question of an appropriate measure \cite{Arzano:2014jfa},\cite{Amelino-Camelia:2013cfa},\cite{Arzano:2015gda}  used to evaluate the integrals defining the heat kernel traces studied here
 is something that first comes into mind.
Indeed, an additional geometric construction,
as revealed through a dual momentum space,  inevitably imposes  some supplementary requirements regarding the measure.
 For instance, borrowing from the just described  example of Snyder and $\kappa-$deformed space, the additional structure that
involves  de Sitter geometry of the corresponding dual curved momentum space may
somewhat  restrict the form of the measure \cite{Arzano:2014jfa}. The restriction itself can be achieved simply by
performing the integration over the relevant group manifold which in this case  is a half of de Sitter space. Some other requirements are also possible, like insisting on the hermiticity of the  coordinate and momentum operators on the  Hilbert space of functions of momenta and thus they may also be considered as a key to find the appropriate measure \cite{preparation}.


\section{Vacuum density in the context of LIV models}


The situation becomes more interesting if we analyse the field operators (\ref{bicrossoperator}) and (\ref{lowenergymaguejosmolin}) within the context of LIV theories.
 Strictly speaking, the required field operators   emerge  respectively  from the dispersions (\ref{bicrossproductbasis-expand}) and  (\ref{mscasimir}), which can be considered as the special cases  of an
exact dispersion relation derived within a particular class of LIV theories \cite{Jacobson:2005bg},\cite{Myers:2003fd}. The latter entails the explicit Lorentz invariance breaking by a certain class of five-dimensional operators added to the effective Lagrangian.

With the dispersions (\ref{bicrossproductbasis-expand}) and (\ref{mscasimir})   being the special cases of an exact dispersion relation, the condition $~M_{UV} << \kappa, ~$ which had to be satisfied before, i.e. in the case of  DSR class of models,  has no relevance anymore in the present context.
Therefore, we  get back to the results (\ref{bicrossoperator-traceheatkernel2})  and (\ref{maguejosmolin-traceheatkernel12})  to  reexamine them and take the limit  $s \kappa^2   << 1.$ One should note that  this limit was forbidden before, however  now  is allowed. It  happens  to be so since in the new circumstances provided by the LIV framework, there is nothing that is incompatible with this limit.
Therefore, we are now allowed to
expand the expressions (\ref{bicrossoperator-traceheatkernel2}) and (\ref{maguejosmolin-traceheatkernel12})  in powers of $\kappa^2 s.$ We first do that for the heat kernel trace (\ref{bicrossoperator-traceheatkernel2}), corresponding to the dispersion (\ref{bicrossproductbasis-expand}), to get
\begin{eqnarray} \label{bicrossoperator-traceheatkernel3}
  \mbox{Tr} \; e^{-s \Box_{\kappa}} \; & = & \; \frac{\Delta \tau V}{{(4\pi)}^{d/2}} \frac{1}{\Gamma(\frac{d-1}{2})} \frac{2}{\sqrt{s}}
    {\bigg(\frac{s}{\kappa^2} \bigg)}^{-\frac{d}{4} }
     \bigg[ {(\sqrt{2})}^{d-5} {\bigg( \frac{1}{\kappa^2} \bigg) }^{\frac{1}{4}}
      \Gamma(\frac{d-1}{4}) s^{\frac{1}{4}}
      - \dfrac{ {(\sqrt{2})}^{d-3}  \Gamma(\frac{d+1}{4}) }{{ \bigg( \frac{1}{\kappa^2} \bigg) }^{\frac{1}{4}}} s^{\frac{3}{4}}  \nonumber  \\
     & + & \dfrac{ {(\sqrt{2})}^{d-7} (d-1) \Gamma(\frac{d-1}{4}) }{{\bigg( \frac{1}{\kappa^2} \bigg) }^{\frac{3}{4}}} s^{\frac{5}{4}}
      -  \frac{1}{3}{(\sqrt{2})}^{d-5} {\bigg( \frac{1}{\kappa^2} \bigg)}^{\frac{3}{4}} \kappa^{4} (d+1) \Gamma(\frac{d+1}{4}) s^{\frac{7}{4}} + \cdots  \bigg].
\end{eqnarray}
From the effective action $W_{eff} [1/\kappa ],$
\begin{equation}  \label{effectiveaction1}
  W_{eff} [1/ \kappa] = -\frac{1}{2} \int_{\epsilon^{2}}^{\infty} \frac{ds}{s}
    \;\mbox{Tr} \; e^{-s \Box_{\kappa}} \; \equiv \; -\Lambda (\epsilon) \Delta \tau V,
\end{equation}
one gets for the energy of the quantum vacuum (note that in the above expression we have introduced the
 quantity   $\Lambda(\epsilon),$ which is the vacuum energy density)
\begin{eqnarray} \label{bicrossoperator-effectiveaction}
   W_{eff} [1/ \kappa] \; & = & \;
    -\frac{1}{2} \frac{\Delta \tau V}{{(4\pi)}^{d/2}} \frac{2}{\Gamma(\frac{d-1}{2})}
    \bigg[ {(\sqrt{2})}^{d-5}  \Gamma(\frac{d-1}{4}) {\kappa}^{\frac{d-1}{2}}
      \dfrac{4}{(d+1) {\epsilon}^{\frac{d+1}{2}}}
          -  {(\sqrt{2})}^{d-3}  \Gamma(\frac{d+1}{4}) {\kappa}^{\frac{d+1}{2}}
       \dfrac{4}{(d-1) {\epsilon}^{\frac{d-1}{2}}}         \nonumber  \\
   & + & {(\sqrt{2})}^{d-7} (d-1) \Gamma(\frac{d-1}{4}) {\kappa}^{\dfrac{d+3}{2}}   \dfrac{4}{(d-3) {\epsilon}^{\frac{d-3}{2}}}
          -  {(\sqrt{2})}^{d-5} \frac{d+1}{3}\Gamma(\frac{d+1}{4}) {\kappa}^{\frac{d+5}{2}}      \dfrac{4}{(d-5) {\epsilon}^{\frac{d-5}{2}}}   + \cdots  \bigg].   \nonumber
\end{eqnarray}
This result is valid for $d > 5.$
 It is interesting to take a look at the case $d=4.$ It gives
\begin{eqnarray} \label{bicrossoperator-effectiveaction1}
  \Lambda (\epsilon) \; & = & \;
   \frac{1}{{(4\pi)}^{2}} \frac{1}{\Gamma(\frac{3}{2})} \;
    \bigg[ \dfrac{4 \; \Gamma(\frac{3}{4})}{5 \sqrt{2} }  \dfrac{{\kappa}^{\frac{3}{2}}}{{\epsilon}^{\frac{5}{2}}}
    - \dfrac{4 \sqrt{2} \; \Gamma(\frac{5}{4})}{3}  \dfrac{{\kappa}^{\frac{5}{2}}}{{\epsilon}^{\frac{3}{2}}}
    + \dfrac{6 \; \Gamma(\frac{3}{4})}{\sqrt{2}}  \dfrac{{\kappa}^{\frac{7}{2}}}{{\epsilon}^{\frac{1}{2}}}
    + finite  \bigg].
\end{eqnarray}
Characteristic feature of the present model is the presence of the logarithmic divergences for the
spaces whose dimension is an odd number. In this respect it is interesting to consider the case with $d=3,$
\begin{eqnarray} \label{bicrossoperator-effectiveaction2}
  \Lambda (\epsilon) \; & = & \;
   \frac{1}{{(4\pi)}^{\frac{3}{2}}}  \;
    \bigg[ \dfrac{ \sqrt{\pi}}{ 2 }  \dfrac{\kappa}{{\epsilon}^{2}}
    - 2 \dfrac{{\kappa}^2 }{\epsilon} + \dfrac{ \sqrt{\pi}}{ 2 }
       {\kappa}^3 \bigg( \ln \Lambda_{IR}  - 2 \ln \epsilon  \bigg) + finite \bigg],
\end{eqnarray}
 where  the constant  $\Lambda_{IR}$ is an appropriately chosen infrared cutoff. Writing only divergent terms, this yields
\begin{eqnarray} \label{bicrossoperator-effectiveaction2}
  \Lambda (\epsilon) \; & = & \;
   \frac{1}{{(4\pi)}^{\frac{3}{2}}}  \;
    \bigg[ \dfrac{ \sqrt{\pi}}{ 2 }  \dfrac{\kappa}{{\epsilon}^{2}}
    - 2 \dfrac{{\kappa}^2 }{\epsilon} - \sqrt{\pi}
       {\kappa}^3   \ln \frac{\kappa \epsilon}{2}   + finite \bigg].
\end{eqnarray}
It is evident that the divergencies in both cases that were presented here  are  milder than in a standard theory with the absence of explicit Lorentz violation.

In the case of the LIV model with the dispersion (\ref{mscasimir}), we turn back to the expression (\ref{maguejosmolin-traceheatkernel12}), which as explained in the Appendix A, is valid only when the dimension $d$ is an odd number.
Following this, the limit $s \kappa^2  << 1$ is taken in (\ref{maguejosmolin-traceheatkernel12}) and the result in this regime is displayed for the cases of $d=3$ and $d=5.$
Thus, for $d=3$ and taking into account (\ref{appendix6}), one readily obtains 
 the result
\begin{equation} \label{appendix15}
   \Lambda (\epsilon) = \frac{\kappa}{16 \pi} \bigg[  \frac{4}{3} \frac{1}{\epsilon^2} + \frac{98\kappa^2}{189} \ln \frac{\Lambda_{IR}}{\epsilon^2} +  finite  \bigg].
\end{equation}
In a similar way, for $d=5$ and with the help of  (\ref{appendix7}), one gets
\begin{equation} \label{appendix16}
   \Lambda (\epsilon) = \frac{\kappa}{64 \pi^2} \bigg[  \frac{1}{18} \frac{1}{\epsilon^4} + \frac{95}{324}
   \frac{\kappa^2}{\epsilon^2}  + \frac{179}{810} \kappa^4 \ln \frac{\Lambda_{IR}}{\epsilon^2} +  finite  \bigg].
\end{equation}
Both these results represent a significant improvement in the UV behaviour of the vacuum energy, when compared to the standard result which predicts  $\Lambda (\epsilon) \sim 1/\epsilon^3$ for $d=3$ and  $\Lambda (\epsilon) \sim 1/\epsilon^5$ for $d=5,$ respectively. In the LIV based models the  limit that was used in calculations ($s \kappa^2 << 1 $) suggests that the UV cutoff $~M_{UV} \sim 1/\epsilon ~$ is significantly above the Lorentz breaking scale $\kappa$,  namely $~M_{UV} >> \kappa, ~$ which itself may typically  be taken to be around UHECR scale or above.

For $d$ even, the function $f(\omega),$ relevant for the analysis carried here (see the formula (\ref{appendix4}) in the Appendix A), is a multivalued function and one needs to be more  cautious,
especially when choosing
 an appropriate contour of integration which needs to  avoid a cut specifying the points in the complex plane where the function is not well defined. We do not treat this case here.



\section{Conclusion}
     We have investigated the UV behaviour of the cosmological constant in the noncommutative geometry background represented by some of the recently most promulgated noncommutative models, including the Majid-Ruegg model based on the bicrossproduct field operator, as well as the Magueijo-Smolin and Snyder model. We have also analysed the field operators that appear within the context of theories with explicit Lorentz violation (LIV). While considerable improvement in the UV behaviour has been found for the vacuum energy in LIV theories, the models based on noncommutative structure of spacetime and DSR frame have shown no improvement at all.
The same conclusion is found for the field operator pertaining to the Majid-Ruegg model   in the  ultrarelativistic regime $\omega >> \kappa$ (see the Appendix B). However, it may be too early to come to such conclusion. Indeed,  with being deeply rooted
in the setting which anticipates a grain-like structure of spacetime, the noncommutative models impose a natural bound on the UV cutoff. Specifically, the UV regulator $\epsilon$ should not be smaller than the Planck length.
Since this makes the limit $\epsilon$ goes to zero unlikely to implement in a full extent,  the vacuum energy 
 in the models considered here turns to be finite,
in accordance with some earlier results \cite{garatini},\cite{Garattini:2011kp}.
Specifically, by considering a modification of the Liouville measure that was induced by noncommutativity, in reference  \cite{garatini} 
the finite contribution to the vacuum energy has  been obtained by considering the quantum fluctuations of the pure gravitational field and in  \cite{Garattini:2011kp} the same result has been obtained in the context of modified dispersion relations.
In the latter paper the modified dispersion relations   enter the formalism
by  indirectly influencing the form of the actual metric, which now becomes energy dependent, leading to the so called
Gravity's Rainbow \cite{Magueijo:2002xx}.  
 Although  the references \cite{garatini},\cite{Garattini:2011kp} consider contributions to vacuum energy coming from gravitational fluctuations  and here the vacuum energy has an origin in the fluctuations of the scalar field, two respective results
  for the vacuum energy density nevertheless share the similar structure, if  according to the previous
 observations, the UV regulator $\epsilon$ is  bound by the natural cutoff $1/M_{Pl} $ or $1/\kappa.$  

 Finally, it is worthy to note that it was also  suggested that due to an intrinsic geometry and symmetry structure of DSR models, some additional care may be required, particularly regarding the measure of integration. This might lead to somewhat different result and we plan to investigate this issue in the forthcoming paper \cite{preparation}.


\noindent{\bf Acknowledgment}\\
I am thankful to S. Mignemi for  fruitful discussions and critical observations.
I also thank G. Rosati for helpful discussion.
The research leading to these results was partly done during the funding from the European Union
Seventh Framework Programme (FP7 2007-2013) under grant agreement n 291823 Marie
Curie FP7-PEOPLE-2011-COFUND NEWFELPRO as a part of a project NCGGBH
which has received funding through NEWFELPRO project under grant agreement n 63.
The partial support by Croatian Science Foundation under the
project (IP-2014-09-9582), as well as the partial financial support by the H2020 CSA Twinning project No. 692194, RBI-T-WINNING
is also acknowledged.

\appendix
\section{Evaluation of the heat kernel trace, the expression (\ref{maguejosmolin-traceheatkernel})}

In this Appendix we evaluate the trace of the heat kernel appearing in the expression (\ref{maguejosmolin-traceheatkernel}).
The aim of the calculation presented here is to obtain a reliable estimation which would retain
all essential features of the heat kernel trace, enabling us to deduce the actual UV structure of the vacuum energy. With this objective in mind,
we first do the $p$ integration, including the angular part, which can be dealt with in the same way as in 
(\ref{bicrossoperator-traceheatkernel}). This leads to
\begin{eqnarray} \label{appendix1}
 \mbox{Tr} \; K(s) \; & = &\; 
    \frac{\Delta \tau V}{{(2\pi)}^{d}} \;
     \int_{-\infty}^{\infty} d\omega  
  \;  e^{-s \omega^2 } 
  e^{ i \frac{2s}{\kappa} {\omega}^3 } 
  \int  d^{d-1}p \;  e^{-s \bigg(1- \frac{2i}{\kappa} \omega \bigg){\bf{p}}^2} \nonumber \\
    & = & \; \frac{\Delta \tau V}{2^d \pi^{d/2}} \;
     \frac{1}{\sqrt{\pi}}  \frac{1}{s^{\frac{d-1}{2}}} \int_{-\infty}^{\infty} 
    \; \frac{e^{-s \omega^2 } e^{i \frac{2s}{\kappa} {\omega}^3 }}{{\bigg( 1 - \frac{2i}{\kappa} \omega     \bigg)}^{\frac{d-1}{2}}} 
     d\omega.   \nonumber
\end{eqnarray}
The $\omega$ integration can be done for $d$ odd by using the residue theorem and taking the contour as in figure 1,
\begin{equation} \label{appendix2}
   \oint_{\Gamma} \frac{e^{-s \omega^2 } e^{i \frac{2s}{\kappa} {\omega}^3 }}{{\bigg( 1 - \frac{2i}{\kappa} \omega     \bigg)}^{\frac{d-1}{2}}} d\omega = \int_{R}^{-R}  \frac{e^{-s \omega^2 } e^{i \frac{2s}{\kappa} {\omega}^3 }}{{\bigg( 1 - \frac{2i}{\kappa} \omega     \bigg)}^{\frac{d-1}{2}}} 
     d\omega + \int_{\gamma(R)} \frac{e^{-s \omega^2 } e^{i \frac{2s}{\kappa} {\omega}^3 }}{{\bigg( 1 - \frac{2i}{\kappa} \omega     \bigg)}^{\frac{d-1}{2}}} d\omega.   
\end{equation}
The contour in figure 1 has been chosen in such a way that it  encircles the pole at $\omega = -i \frac{\kappa}{2}$ of the
function under the integral and  that upon integration it is traced   in a counterclockwise direction.
While the integral over the closed contour is determined by the residuum of the function $f(\omega)$ at the pole, 
\begin{equation} \label{appendix3}
   \oint_{\Gamma} \frac{e^{-s \omega^2 } e^{i \frac{2s}{\kappa} {\omega}^3 }}{{\bigg( 1 - \frac{2i}{\kappa} \omega     \bigg)}^{\frac{d-1}{2}}} d\omega = 2\pi i ~ {\mbox{Res}}_{\omega = -i \frac{\kappa}{2}} f(\omega),
\end{equation}
the integral over the semicircle of radius
$R$ needs some additional care. In particular, we want to investigate its behaviour as $R \longrightarrow \infty.$ 
Note that  the presence of an additional exponential factor in the numerator precludes a straighforward application of the Jordan's lemma.
Nevertheless, even if the required integral  doesn't vanish in the  limit $R \longrightarrow \infty$, it is
of interest to find an analytic expression which approximates its value and at the same time
contains the  features essential for deducing the UV structure of $\Lambda(\epsilon)$.

\begin{figure}
\begin{center}
 \includegraphics[scale=0.8]{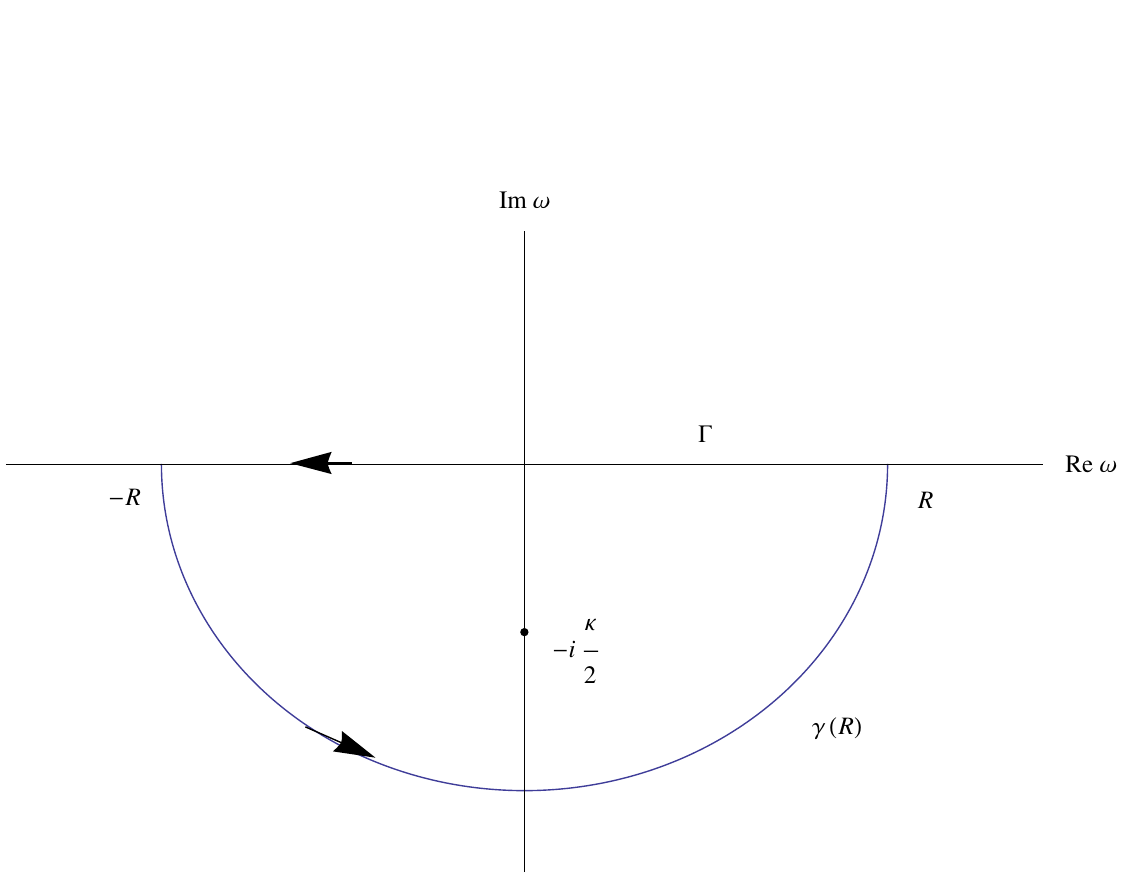}  \\ Fig.1
\end{center}
\end{figure}

In order to find a residue of the function
\begin{equation} \label{appendix4}
  f(\omega) = \frac{e^{-s \omega^2 } e^{i \frac{2s}{\kappa} {\omega}^3 }}{{\bigg( 1 - \frac{2i}{\kappa} \omega     \bigg)}^{\frac{d-1}{2}}},
\end{equation}
at the pole $\omega = -i \frac{\kappa}{2}$ for $d$ odd, we write the appropriate Laurent series around this point
\begin{eqnarray} \label{appendix5}
  f(\omega)  &= & \frac{1}{{\bigg( - \frac{2i}{\kappa} \bigg)}^{\frac{d-1}{2}}} \Bigg[ \frac{1}{{\bigg( \omega + i\frac{\kappa}{2} \bigg)}^{\frac{d-1}{2}}}  - \frac{1}{2} is \kappa \frac{1}{{\bigg( \omega + i\frac{\kappa}{2} \bigg)}^{\frac{d-3}{2}}}  \nonumber \\
 &+ & \big( 4s - \frac{1}{8} s^2 \kappa^2  \big) \frac{1}{{\bigg( \omega + i\frac{\kappa}{2} \bigg)}^{\frac{d-5}{2}}} + \frac{i\big( \frac{2s}{\kappa} - 4s^2 \kappa + \frac{1}{48} s^3 \kappa^3  \big)}{{\bigg( \omega + i\frac{\kappa}{2} \bigg)}^{\frac{d-7}{2}}} 
   + {\mathcal{O}}{(\omega + i \frac{\kappa}{2})}^{-\frac{d-9}{2}} 
 \Bigg],
\end{eqnarray}
giving residues of  $f$ for $d=3,5,7,9,$ respectively. For the purpose of illustrating what happens to the cosmological constant  in the model with Casimir (\ref{mscasimir}),
we  focus on two special cases when the number of dimensions is $d=3$ and $d=5.$ The residues in these two  cases can be   respectively   read out from the series (\ref{appendix5}) as
\begin{equation} \label{appendix6}
 {\mbox{Res}}_{\omega = -i \frac{\kappa}{2}} f(\omega) = i\frac{\kappa}{2}, \quad {\mbox{for}} \; d=3,
\end{equation}
\begin{equation} \label{appendix7}
 {\mbox{Res}}_{\omega = -i \frac{\kappa}{2}} f(\omega) = i s \frac{\kappa^3}{8}, \quad {\mbox{for}} \; d=5.
\end{equation}
We next turn to the contour integral over the semicircle of radius $R$ and estimate its absolute value $I_R$, which will
serve to put the upper bound on the integral itself. Hence,
\begin{equation} \label{appendix8}
  I_R  =   \Bigg| \int_{\gamma(R)} \frac{e^{-s \omega^2 } e^{i \frac{2s}{\kappa} {\omega}^3 }}{{\bigg( 1 - \frac{2i}{\kappa} \omega     \bigg)}^{\frac{d-1}{2}}} d\omega  \Bigg| 
  \equiv  \Bigg| \int_{\gamma(R)} g(\omega) ~e^{i \frac{2s}{\kappa} {\omega}^3 } d\omega  \Bigg| 
  = \Bigg| \int_{\gamma(R)} g(R e^{i\theta}) ~e^{i \frac{2s}{\kappa} R^3 e^{3i\theta}} iR e^{i\theta}d\theta  \Bigg|, \nonumber 
\end{equation}
where the function $g$ is introduced as $g(\omega) =   e^{-s \omega^2 } \bigg/{\bigg( 1 - \frac{2i}{\kappa} \omega     \bigg)}^{\frac{d-1}{2}} $.

Successively we have
\begin{equation} \label{appendix9}
  I_R  =  R ~\Bigg| \int_{-\pi}^{0} g(R e^{i\theta})
e^{i \frac{2s}{\kappa} R^3 ( \cos 3\theta + i \sin 3\theta) } i e^{i\theta} d\theta  \Bigg| 
  \le R   \int_{-\pi}^{0} \bigg| g(R e^{i\theta}) \bigg| ~e^{-\frac{2s}{\kappa} R^3 \sin 3\theta} d\theta  
    \le  RM_R  \int_{-\pi}^{0}  ~e^{-\frac{2s}{\kappa} R^3 \sin 3\theta} d\theta,  \nonumber \\
\end{equation}
where $$ M_R = \max_{\theta \in [-\pi, 0 ]} \bigg|  g(R e^{i\theta}) \bigg| $$
is the maximal value of $g$ over the whole semicircle.

\begin{figure}
\begin{center}
\includegraphics[width=9cm]{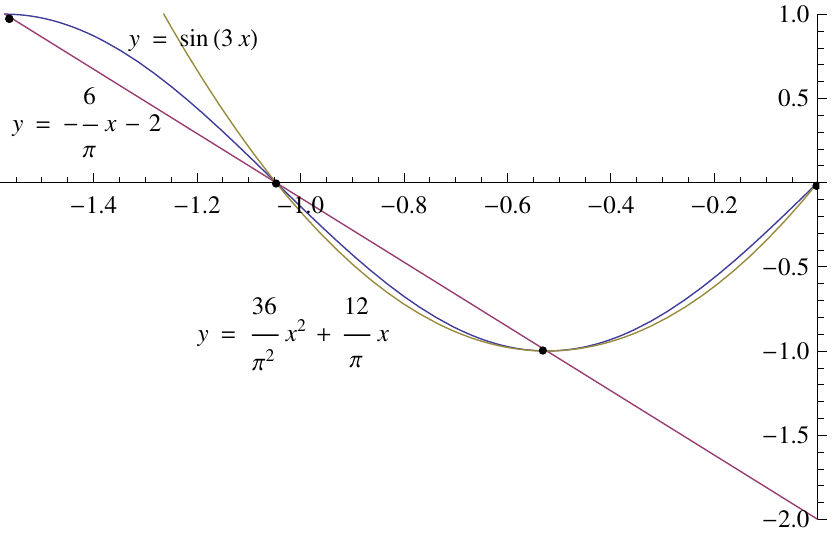}  \\ Fig.2
\end{center}
\end{figure}

As can be seen from  figure 2, the interval $[-\pi, 0]$
can be divided into 6 subintervals, which can be analysed separately to give
\begin{eqnarray} \label{appendix10}
  I_R  &\le &  RM_R ~\Bigg[ 2 \int_{-\pi/2}^{-\pi/3} 
   ~e^{-\frac{2s}{\kappa} R^3 \sin 3\theta} d\theta + 4\int_{-\pi/6}^{0} 
   ~e^{-\frac{2s}{\kappa} R^3 \sin 3\theta} d\theta        \Bigg]  \nonumber \\
 & \le& RM_R ~\Bigg[ 2 \int_{-\pi/2}^{-\pi/3} 
   ~e^{-\frac{2s}{\kappa} R^3 \bigg( -\frac{6\theta}{\pi} -2  \bigg) } d\theta + 4\int_{-\pi/6}^{0} 
   ~e^{-\frac{2s}{\kappa} R^3 \bigg( \frac{36}{\pi^2} \theta^2 +\frac{12}{\pi} \theta \bigg) } d\theta      \Bigg]. 
\end{eqnarray}
The second line is justified from figure 2. Namely, as is readily  seen from there, on the interval $[-\pi/2, -\pi/3]$  the straight line
 can be drawn through the points $(-\pi/2, 1)$ and
$(-\pi/3, 0)$
 which lies below the curve $\sin 3\theta \;  .$
  In the same way, on the interval $[-\pi/3, 0]$ the parabola  can be drawn
 which passes through the   points
$(-\pi/3, 0), ~(-\pi/6, -1)$ and $(0, 0)$, and which  lies below the  curve $\sin 3\theta$ in the considered interval. While the straight line has the equation $y = -\frac{6\theta}{\pi} -2, $ the parabola is determined by $y= \frac{36}{\pi^2} \theta^2 +\frac{12}{\pi} \theta$. This means that on the interval $[-\pi/2, -\pi/3]$ 
$$
  \sin 3\theta \ge -\frac{6}{\pi} \theta -2,  \quad \mbox{implying}  \quad e^{-\frac{2s}{\kappa} R^3 \sin 3\theta} \le e^{-\frac{2s}{\kappa} R^3 \bigg( -\frac{6\theta}{\pi} -2  \bigg) }
$$
and on the interval $[-\pi/6, 0]$
$$
  \sin 3\theta \ge \frac{36}{\pi^2} \theta^2 +\frac{12}{\pi} \theta,  \quad \mbox{implying}  \quad e^{-\frac{2s}{\kappa} R^3 \sin 3\theta} \le e^{-\frac{2s}{\kappa} R^3 \bigg( \frac{36}{\pi^2} \theta^2 +\frac{12}{\pi} \theta \bigg) }.
$$
This finally gives
\begin{eqnarray} \label{appendix11}
  I_R  &\le &  RM_R ~\Bigg[ 2 \frac{\pi}{12s} \frac{\kappa}{R^3} \bigg(1- 
   ~e^{-\frac{2s}{\kappa} R^3 } \bigg) + 4 \frac{\pi^{3/2}}{12} \sqrt{\frac{\kappa}{2s}}
    \frac{e^{\frac{2s}{\kappa} R^3 }}{\sqrt{R^3}}      
  {\mbox{erf}} \bigg(\sqrt{\frac{2s}{\kappa} R^3} \bigg) \Bigg]      \nonumber \\
 & =& \frac{e^{sR^2}}{{\bigg( 1+ \frac{4}{\kappa} R   + \frac{4}{\kappa^2} R^2 \bigg)}^{\frac{d-1}{4}}}
   ~\Bigg[  \frac{\pi}{6s} \frac{\kappa}{R^2} \bigg(1- 
   ~e^{-\frac{2s}{\kappa} R^3 } \bigg) +  \frac{\pi^{3/2}}{3} \sqrt{\frac{\kappa}{2s}}
    \frac{e^{\frac{2s}{\kappa} R^3 }}{\sqrt{R}}      
  {\mbox{erf}} \bigg(\sqrt{\frac{2s}{\kappa} R^3} \bigg) \Bigg],
\end{eqnarray}
where in the last line  $M_R$ has been explicitly evaluated. The expression on the right hand side obviously diverges in the limit $R \rightarrow \infty $ and consequently the contour integral  over the semicircle $\gamma(R)$ does not vanish in that limit. However,
 instead of requiring $R \rightarrow \infty, $ we can   to a reasonable approximation take the semicircle radius $R$ to be of the order of the  LIV or NC scale $\kappa,$
 $~R \sim \kappa.$ This assertion is further supported by the very fact that in the Magueijo-Smolin model the maximal energy is neither more, nor less than $\kappa$.
 This puts the following upper bound on the integral $I_R,$
\begin{equation} \label{appendix12}
  I_R  ~ \le ~ \frac{e^{s\kappa^2}}{9^{\frac{d-1}{4}}} \Bigg[  \frac{\pi}{6\kappa s} \bigg(1- e^{-2s \kappa^2}   \bigg)   + \frac{\pi^{3/2}}{3} \frac{e^{2s\kappa^2}}{\sqrt{2s}} {\mbox{erf}} \bigg(\sqrt{2s \kappa^2} \bigg) \Bigg].
\end{equation} 
As a next step, the contour integral over $\gamma (R)$ is approximated by the expression on the right hand side 
of eq.(\ref{appendix12}), giving rise to the heat kernel trace of the form
\begin{eqnarray} \label{appendix13}
 \mbox{Tr} \; K(s) \; & = &\; 
     \frac{\Delta \tau V}{2^d \pi^{d/2}} \;
     \frac{1}{\sqrt{\pi}}  \frac{1}{s^{\frac{d-1}{2}}} \int_{-\kappa}^{\kappa} 
    \; \frac{e^{-s \omega^2 } e^{i \frac{2s}{\kappa} {\omega}^3 }}{{\bigg( 1 - \frac{2i}{\kappa} \omega     \bigg)}^{\frac{d-1}{2}}}  d\omega   \nonumber \\
  & = & \; \frac{\Delta \tau V}{2^d \pi^{d/2}} \;
     \frac{1}{\sqrt{\pi}}  \frac{1}{s^{\frac{d-1}{2}}} \Bigg(  - \oint_{\Gamma}
    \; \frac{e^{-s \omega^2 } e^{i \frac{2s}{\kappa} {\omega}^3 }}{{\bigg( 1 - \frac{2i}{\kappa} \omega      \bigg)}^{\frac{d-1}{2}}}  d\omega   + \int_{\gamma (R)}
    \; \frac{e^{-s \omega^2 } e^{i \frac{2s}{\kappa} {\omega}^3 }}{{\bigg( 1 - \frac{2i}{\kappa} \omega     \bigg)}^{\frac{d-1}{2}}}  d\omega  \Bigg).
\end{eqnarray}
This in turn gives rise to the effective action
\begin{eqnarray} \label{appendix14}
    W_{eff} [1/ \kappa] \; & = & \;  -\frac{1}{2} \int_{\epsilon^2}^{\infty} \frac{ds}{s} ~\mbox{Tr} \; K(s)  \nonumber \\
   & = & \; -\frac{1}{2} \int_{\epsilon^2}^{\infty} \frac{ds}{s} ~ \frac{\Delta \tau V}{2^d \pi^{d/2}} \;
     \frac{1}{\sqrt{\pi}}  \frac{1}{s^{\frac{d-1}{2}}} \Bigg(  -  2\pi i ~ {\mbox{Res}}_{\omega = -i \frac{\kappa}{2}} f(\omega)  \nonumber \\  
  &+&  \frac{e^{s\kappa^2}}{9^{\frac{d-1}{4}}} \Bigg[  \frac{\pi}{6\kappa s} \bigg(1- e^{-2s \kappa^2}   \bigg)   + \frac{\pi^{3/2}}{3} \frac{e^{2s\kappa^2}}{\sqrt{2s}} {\mbox{erf}} \bigg(\sqrt{2s \kappa^2} \bigg) \Bigg]  \Bigg),
\end{eqnarray}
with $~f(\omega) ~$  as in (\ref{appendix4}). This result forms a basis for the analysis carried in both, in section III as well as in section IV.

\section{Ultrarelativistic limit for bicrossproduct basis}

It is interesting to analyse a UV behaviour of the field operator (\ref{bicrossproductbasis}) in the ultrarelativistic regime with an energy $p_0$ being relatively greater than the NC (or LIV) scale $\kappa$ 
 (i.e. $p_0$ is considered to be sufficiently greater than $\kappa$ so that ~ $\exp{(p_0/\kappa)} >> 1$).
Therefore, in this appendix we
 analyse a divergency structure of the vacuum energy in the ultrarelativistic limit. For that purpose, we consider the Casimir 
 (\ref{bicrossproductbasis}). In the limit of ultrahigh energies (where $p_0 $ is relatively greater than $\kappa$ and $ ~\exp{(p_0/\kappa)} >> 1 $), the Casimir (\ref{bicrossproductbasis})
  reduces to
\begin{equation}
  -m^2 = M^2 = {( {\bf{p}}^2 - \kappa^2)} e^{-i \frac{p_d}{\kappa}},
\end{equation}
leading to the Laplacian
\begin{equation}
  \hat{O} \equiv \Box_{\kappa}= {( - \partial_i \partial_i - \kappa^2)} e^{- \frac{1}{\kappa} \frac{\partial}{\partial \tau}}.
\end{equation}
Note that  the energy   $p_0$, although spreading over  the ultrarelativistic range, may still fall well below the Planck scale $M_{Pl}$ if the NC (or LIV) scale is somewhere around the UHECR (ultra-high-energy cosmic rays) energies.  
Applying the similar steps as before, the relevant heat kernel trace appears to be given by
\begin{eqnarray} \label{ultrabicross-traceheatkernel}
  \mbox{Tr} \; K(s) \; = \frac{\Delta \tau V}{{(2\pi)}^{d}} \;
    \frac{2 {\pi}^{\frac{d-1}{2}}}{\Gamma(\frac{d-1}{2})}
     \int_{-\infty}^{\infty} d\omega
      \int_{0}^{\infty}  {|{\bf{p}}|}^{d-2} e^{- s e^{-i\frac{\omega}{\kappa}} ({\bf{p}}^2 - \kappa^2) }
      d {|{\bf{p}}|}.
\end{eqnarray}
After making the integration over momentum, one arrives at
\begin{eqnarray} \label{ultrabicross-traceheatkernel1}
  \mbox{Tr} \; K(s) \; = \frac{\Delta \tau V}{{2}^{d} {\pi }^{\frac{d+1}{2}}} \;
     s^{\frac{1-d}{2}}  \int_{-\infty}^{\infty}
        e^{-i  \frac{\omega}{\kappa} \frac{1-d}{2}} \; e^{ s\kappa^2 e^{-i  \frac{\omega}{\kappa}}}
     \; d\omega,
\end{eqnarray}
which in turn may be written as
\begin{eqnarray} \label{ultrabicross-traceheatkernel2}
  \mbox{Tr} \; K(s) \; =  \frac{\Delta \tau V}{{2}^{d} {\pi }^{\frac{d+1}{2}}} \;
     s^{\frac{1-d}{2}}  \int_{-\infty}^{\infty}
        e^{-i  \frac{\omega}{\kappa} \frac{1-d}{2}} \; e^{ -is\kappa^2 \sin{\frac{\omega}{\kappa}}}
       e^{ s\kappa^2 \cos{\frac{\omega}{\kappa}}} \; d\omega.
\end{eqnarray}
A brief inspection of the above relation by using a saddle point method gives for the dominant contribution a term
whose dependence on  the small adiabatic parameter $s$ goes as $\sim \frac{e^{s \kappa^2}}{ s^{d/2}},$
which shows no improvement in the UV behaviour of $\Lambda (\epsilon),$ when compared to the standard theory.
Indeed, the additional exponential factor, which is absent in the standard theory, might only have an impact in the IR regime, but cannot regularize the vacuum energy density in the UV regime.


\end{document}